%% file: LRR_MO.tex
\documentclass[sigconf]{acmart}
\renewcommand\footnotetextcopyrightpermission[1]{} 
\DeclareMathOperator*{\E}{E}

\DeclareMathOperator*{\I}{I}
\DeclareMathOperator*{\T}{T}

\setcopyright{rightsretained}
\acmDOI{10.475/123_4}

\acmISBN{123-4567-24-567/08/06}

\acmConference[RecSys'17]{ACM RecSys Conference}{August 2017}{Como, Italy} 
\acmYear{2017}
\copyrightyear{2017}

\begin{document}
% Title portion. Note the short title for running heads 
\title[]{A Multi-Objective Learning to re-Rank Approach to Optimize Online Marketplaces for Multiple Stakeholders}  
%\title[]{A Multi-Objective Learning to re-Rank Approach to Optimize Customer Preferences and MarketPlace Compensation}  
%\title[]{A Learning to re-Rank Approach for Multi-Stakeholder Recommendations}  
\author{Phong Nguyen}
\orcid{1234-5678-9012-3456}
\author{John Dines}
%\author{Ioannis Partalas}
\author{Jan Krasnodebski}
\affiliation{%
  \institution{Expedia LPS}
  \streetaddress{42-44 Rue du XXXI Decembre}
  %\city{Geneva}
  \state{Geneva}
  %\postcode{23185}
  \country{Switzerland}}

\input{abstract.tex}





\thanks{}

\maketitle

\renewcommand{\shortauthors}{P. Nguyen, J. Dines \& J. Krasnodebski}

\input{introduction.tex}

\input{related.tex}

\input{LR.tex}

\input{problem_setting.tex}

\input{proposed_approach.tex}

\input{experiment.tex}

\input{conclusion.tex}

\bibliographystyle{ACM-Reference-Format}
\bibliography{ref,joachims}

\end{document}

%% file: abstract.tex
\begin{abstract}
Multi-objective recommender systems address the difficult task of recommending items that are relevant to multiple, possibly conflicting, criteria.  However these systems are most often  designed to address the objective of one single stakeholder, typically, in online commerce, the consumers whose input and purchasing decisions ultimately determine the success of the recommendation systems. 
In this work, we address the multi-objective, multi-stakeholder,   recommendation problem involving one or more objective(s) per stakeholder. In addition to the consumer stakeholder, we also consider two other stakeholders; the suppliers who provide the goods and services for sale %and set the prices of those goods
and the intermediary who is responsible for helping connect consumers to suppliers via its recommendation algorithms.
% and who takes a commission from the supplier and/or consumer in exchange for this service. 
We analyze the multi-objective, multi-stakeholder, problem from the point of view of the online marketplace intermediary whose objective is to maximize its commission through its recommender system. We define a multi-objective problem relating all our three stakeholders which we solve with a novel learning-to-re-rank approach  that makes use of a novel regularization function 
based on the Kendall tau correlation metric and its kernel version;  
given an initial ranking of item recommendations built for the consumer, we aim to re-rank it such that the new ranking is also optimized for the secondary objectives while staying close to the initial ranking.  
We evaluate our approach on a real-world dataset of hotel recommendations provided by Expedia where we show the effectiveness of our approach against a business-rules oriented baseline model.   

%Recommender systems are typically designed towards one specific target, the customer preferences which they try to optimize. In this paper, we propose a learning to re-rank approach for multiple objective problem amenable to optimize both the customer and intermediary targets. Given a ranking of items originally optimized for the customer, we learn how to re-rank this first pass ranking so that it is also optimized for the marketplace target while staying close to the original ranking order. Experimental results on a Expedia dataset show the efficiency of our algorithm compared to a simple line search approach.
\end{abstract}

%% file: introduction.tex
\section{Introduction}

In an academic setting, learning to rank problems are typically framed in terms of the optimization of an objective function that orders items from most relevant to least relevant according to a given search query~\cite{furnkranz2003pairwise,furnkranz2010preference}.   Relevance is determined either directly from hand labelled data or more commonly, for large datasets, implicitly via user relevance feedback (e.g. using click logs of a web application)~\cite{Joachims/02c,Joachims/Radlinski/07a,Radlinski/Joachims/05a}.  In a practical setting, such as online commerce, optimizing for relevance is considered synonymous with optimization of conversion rate (CVR) or click-through rate (CTR) of the consumer~\cite{Radlinski/etal/10a,Yue/etal/10a,Radlinski/etal/08b}. However, such application of learning to rank and, more generally recommender systems,  often demands several objectives be taken into account ~\cite{Rodriguez2012,adomavicius2015multi,kang2012learning,dai2011multi,svore2011learning,agarwal2011click}. In addition to the aforementioned relevance criterion, the business may also demand the incorporation of additional factors representing the interests of the various stakeholders, such as product supply, lifetime customer value as well as short-term profit. 
{\let\thefootnote\relax\footnote{Presented at the 2017 Workshop on Value-Aware and Multistakeholder Recommendation.}}

In online marketplaces one can consider three primary stakeholders, 1) first and foremost the {\em consumer} whose input and purchasing decisions ultimately determine the success of the recommendation systems, 2) the {\em suppliers} who provide the goods and services for sale, and 3) the {\em intermediary} who is responsible for helping connect consumers to suppliers via its recommendation algorithms and who %takes a commission from the supplier and/or consumer in exchange 
receive compensation for this service.

Ideally, all these stakeholder interests and factors should then feature in the learning algorithm, but in practice this requires both a precise mathematical specification of each objective as well as a learning algorithm that is able to solve for multiple (and potentially opposing) objectives.   Instead, to avoid such complexity, ad hoc business rules are often applied on top of the underlying recommendation system to the overall detriment of some or all of our optimization criteria. This is not ideal and can impact the business model of the online marketplace through a decrease in the conversion rate of visitors because of sub-optimal recommendations, or the abandonment of suppliers from the marketplace because of lack of sales.  

In this paper we take a novel machine learning perspective on the multi-objective (stakeholder) recommendation problem. 
%by following a two-stage approach in which at first stage an initial ranking is obtained by focusing on customer's preferences while in a second stage a new ranking is built on top of the first such that a second (and potentially third or more) stakeholder's interest is also taken into account while staying as close as possible to the initial ranking order.  
We frame our approach as a learning-to-re-rank problem in which we define a novel optimization problem based on the kernelized Kendall tau metric~\cite{jiao2016kendall}.
%More specifically, our goal is to get a final recommendation of products that are both good for the customer (in terms of its preference) and for the intermediary (in terms of its commission or margin profit).   
To our knowledge, this is the first time a multi-objective recommendation problem is addressed within the learning-to-rerank framework. Our proposal is also the first to address the difficult problem of optimizig for the profit of the online marketplace while respecting consumer preferences. 

%Most recommender systems are typically designed to satisfy one stackholder's target, the customer preferences, by solving one single objective function, like to recommend movies to users based on the star rating of their previous seen movies. When starting to consider other stackholder's targets like compensations of the movie providers, the problem becomes much more difficult because it involves solving multiple objective functions, one per stackholder, which are most often contradictory between them. For instance, 

%It is well known that in such situation there exists several Pareto optimality solutions 

%E-commerce platforms like Expedia.com have a very challenging task to daily deliver to their customers the best product in terms of quality and price. Finding the right balance between these two components, quality and price, is the most important aspect because it not only impacts the customer choices but also the margins or compensations that the platform can make according to the deal it has with the different suppliers. This involves then solving a more complex recommendation problem than usual in which not only the customer preferences are the primary targets but also the supplier compensations.  

%% file: related.tex
\section{Related Works}

%To our knowledge,we are the first to address the 

Most works on multi-objective recommendation systems do not deal with multiple stakeholders as we do here. They mainly focus on one stakeholder, typically the consumer,  who can have multiple relevance criteria including proximity, reputation, freshness, diversity, engagement and other engagement metrics to only cite a few~\cite{Rodriguez2012,adomavicius2015multi,kang2012learning,dai2011multi,svore2011learning,agarwal2011click}. 
We can distinguish two main approaches in these works; to learn how to aggregate labels in order to build one global model~\citep{adomavicius2015multi,dai2011multi,svore2011learning,agarwal2011click}, or to learn how to aggregate multiple models, one built for each aspect/facet of relevance~\citep{Rodriguez2012,kang2012learning}. Experimenting with the two approaches on a multi-aspect problem, \citep{kang2012learning} found the model aggregation method was more effective than the label aggregation method.

The model aggregation approach~\citep{Rodriguez2012,kang2012learning} is close to our work since it amounts to learning a linear combination of specific-objective models, similarly to what we do here (see section \ref{sec:prob}). 
Also related is the work of \cite{wang2012robust} who propose a multi-objective optimization approach to trade-off the performance of a learning to rank model  between risk and reward against a baseline model. 
The difference between our work and the aforementioned work is that we address the multi-objective problem as a learning-to-re-rank problem. 
%where the second-stage ranking is able to satisfy both first and second stages objectives. 
%More precisely, we solve our multi-objective problem by learning to re-rank an initial ranking such that both first- and second-stage  objectives are satisfied. 
%Our approach is thus also different from previous learning to re-rank approaches since we focus on a multi-objective problem while their goal is to improve an initial ranking but for the same objective~\cite{jain2011learning,kang2011learning}.

Lastly, we should note the work on revenue optimization in sponsored search, in which the dual problem of placing ads relevant to the customer while maximizing revenue of the platform is approached~\cite{he2013game,golovin2009online}. 
These works are however not considering the problem from a recommendation point of view, but from the view of auction mechanisms and game-theory. 
The only exception is the work of \cite{krasnodebski2016}, in which the authors provide a mathematical definition of the relation between suppliers and the Expedia recommendation engine from a revenue optimization perspective.

In this work, we propose to go one step ahead of the initial work of \cite{krasnodebski2016} by defining the multi-stakeholder problem for revenue optimization as a machine learning problem.

%% file: LR.tex
\section{Learning to rank}

Before proceeding to our problem statement, we will first describe the task of learning to rank which forms the basis of the online marketplace's recommendation algorithm. 
%to produce a first stage ranking of items to recommend to the customers.  
In the following, we will denote by $\mathbf{X} \in \mathcal{X}$ the set of $n$ items with consumer preference labels $\mathbf{y} \in \mathcal{Y}$. Also let $f: \mathcal{X} \times \mathcal{Y} \rightarrow \mathbb{R}$ be the recommender system\footnote{In this paper, we let the form of $f$ be unspecified as it depends on the type of model used by the recommender system.} used at the first stage to produce the vector of scores $\mathbf{u} \in \mathbb{R}^n$ (often transformed to probabilities) that describe the product's utility with respect to a consumer search, that is:
\begin{align*}
\mathbf{u} &= \{ \Pr(y_i=1|\mathbf{x}_i) \}_{i=1}^n \\
 &= \{ f(\mathbf{x}_i, y_i) \}_{i=1}^n
\end{align*}

These utility scores are then usually sorted from high to low in order to deliver a ranking of product recommendations to the consumer. To produce the scores state-of-the-art recommender systems~\cite{burges2010ranknet,burges2005learning,burges2011learning,Joachims/02c} typically proceed by following the pairwise or list-wise approaches in which a ranking metric like the Normalized Discount Cumulative Gain (NDCG~\cite{jarvelin2000ir}) is optimized over pairs or lists of consumer preferences so that items that are preferred are ranked in top positions of the search result page. In particular, the pairwise approach LambdaRank~\cite{burges2010ranknet} is one of the most effective approaches~\cite{donmez2009local} due to its ability to minimize a surrogate loss $\mathcal{L}_r$ of the NDCG metric (equivalent to maximizing this metric):
\begin{align}
\label{f:Lr}
\min_f \mathcal{L}_r(\mathbf{y} | \mathbf{X}) = \sum_{y_i \geq y_j} \log(1 + \exp( -\theta (u_i - u_j) )) |\Delta_{ij}^ \text{NDCG}|
\end{align}
where $u_i = f(\mathbf{x}_i, y_i)$ and $\Delta_{ij}^\text{NDCG} = (2^{y_i} - 2^{y_j}) (\log(1+r(u_i)) - \log(1 +r(u_j)))$ is the cost in terms of NDCG of swapping item $i$ with item $j$ from position $r(u_i)$ to position $r(u_j)$.

It is worth noting that the recommender system, $f$, described here is designed to solely focus on the consumer preferences or interest. It does not take into account the other stakeholders' interests, such as supplier profit and intermediary commission, to also maximize the business value of the recommendation. In the following section a detailed problem statement is provided that takes into account multiple stakeholders, which we frame as a multi-objective problem. Our goal is to find a linear combination of the intermediary's and supplier's interests with the consumer preferences provided by our recommender system $f$ such that all three stakeholders are satisfied by the final product recommendations

%% file: problem_setting.tex
\section{Problem Setting}
\label{sec:prob}

By playing the pivot role between consumers and suppliers a successful marketplace is able to maximize its compensation using two primary levers: by maximising the number of transactions on its online platform through its recommender system $f$, which is typically measured by the consumer's conversion rate (CVR) or click through rate (CTR), as well as endeavouring to obtain the maximum per-transaction commission from the suppliers.  A common commission model\footnote{This the model most commonly applied in traditional online travel agencies (OTAs).  Other commission models, for example, where the consumer pays the commission also exist and could equally employ the methodology described here.} has the supplier paying the intermediary a portion of the sales transaction (herein referred to as margin, $m$), which is equal to the selling price $p$ of the product minus the supplier's cost $c$, i.e. $m = p - c,$ and in the expectation it is equal to:
\begin{align}
\label{f:E}
\E[m|\mathcal{X},\mathcal{Y}] &= \E[ \Pr(\mathcal{Y}=1|\mathcal{X}) \times m ] \nonumber \\ 
&= \E[ f(\mathcal{X}, \mathcal{Y}) \times m ]
\end{align}
%where $f$ is the online marketplace's recommender system described before. 

In order to maximize its commission the intermediary has then to solve the following objective function $\mathcal{L}$ given by maximum log-likelihood of (\ref{f:E}):
\begin{align}
\label{f:L}
\max_{\alpha,\beta} \mathcal{L}(\mathbf{m}|\mathbf{u}) &= \sum_{i=1}^n \log(u_i) + \alpha \log(p_i) + \beta \log( m_i / p_i) 
\end{align}
where $\alpha$ and $\beta$ are tuning parameters (originally set to one) that gives the importance of the log margin component which we have separated into two parts, price and margin percent; i.e. $\log(m) = \log(m/p) + \log(p)$. 

On the surface, this problem (\ref{f:L}) resembles a simple linear combination of three factors. In reality there is the potential for interactions between these factors, as such the task more closely resembles  a {\em multi-objective} optimization problem where each factor corresponds respectively to the consumer's interest as defined by the score $u$, the supplier's interest as defined by price $p$ and the intermediary's interest defined here as the percent commission $m/p$. 
In other words, the intermediary has to solve a problem relating our three stakeholders, in particular, the relation between consumers and suppliers, which is by nature contradictory since these two stakeholders typically have opposing preferences in terms of the price of the product of interest. 
Likewise, the commission percent is also a potential point of contention between the intermediary and suppliers since a higher commission benefits one party at the cost to the other, whereas the consumer is little concerned with and is generally not aware of the commission amount. 

The intrinsic relation between our three stakeholders that we have just described represents the core problem of any online marketplace.
For such problems there is no unique Pareto optimal solution because of the aforementioned potentially complex and opposing interactions of the different stakeholder interests.  
Therefore, the importance one can confer to the suppliers and intermediary stakeholders in this relation %(as described by the weights $\alpha$ and $\beta$) 
most often depends on some business rules following a short term profit,  which is not optimal and 
%It is worth noting that without a careful assignment of those weights such approach 
can easily degrade the performance of all the stakeholder objectives 
%the intermediary's recommendation algorithm, as well as to strongly impact the overall profit of both the suppliers and intermediary stakeholders  
in the long term view.

In this paper, we propose to go beyond the typical business approach by learning how to adjust the weight $\beta$ of the intermediary with respect to the input attribute $\mathbf{X}$, the consumer preference utility scores $\mathbf{u}$ and the suppliers' margins $\mathbf{m}$,  that is:
\begin{align}
\label{f:L3}
\max_\beta \mathcal{L}(\mathbf{m}|\mathbf{u},\mathbf{X}) &= \sum_{i=1}^n \log(u_i) + \alpha \log(p_i) + \beta(\mathbf{x}_i,m_i) \log( m_i / p_i)
\end{align} 
where in this work we treat $\alpha$ as a fixed hyper parameter. 

In the following section, we will frame problem (\ref{f:L3}) as a novel learning-to-re-rank problem whose task is to re-rank the recommended items such that the new ranking order provided by (\ref{f:L3}) stays as close as possible to the original ranking order $\mathbf{u}$ while maximizing at the same time the margins $\mathbf{m}$ of the online marketplace.

%% file: proposed_approach.tex
\section{Learning to re-rank for Multi-Objective Problem}

We will now present a novel learning-to-re-rank approach to address the multi-objective problem (\ref{f:L3}). Learning-to-re-rank is the machine learning approach to improve an initial ranking of items by taking into account second-order item information like co-clicks or co-image similarity~\cite{jain2011learning,kang2011learning}. 
%Typical approaches make use of similarity information between pairs of items like co-clicks or co-image similarity to produce a new ranking supposed to deliver better recommendations than the original ranking. 
In this work, we consider the task of promoting items to higher positions that better satisfy a secondary objective function (in this use case, the transaction commission) while staying  as close as possible to the original ranking. 

Intuitively, this amounts to finding pairs of items for which a swap promoting an item with higher compensation will have little impact on the other pairs. 
To do this, we will make use of the Kendall tau correlation metric~\cite{kendall1938new} as this metric works on the level of pairwise agreements of two rankings. 
More precisely, we will operate on the relative pairwise order of the items -- and not their scores -- since our task is to produce a new ranking of items which maximizes the intermediary's compensation while minimizing the distance between the new and old rankings.

%In the following, we will propose a novel optimization problem able to address such task. 
%Even though this can be done naively during inference by looping over all item's pairs and selecting the best candidate pairs to re-rank, there is no guarantee that the original ranking will be preserved and the quadratic complexity of such procedure can be very prohibitive on large set of items. In the following, we will propose instead a novel learning-to-re-rank approach whose application is linear in the number of items and thus amenable to large scale recommender systems.

\subsection{Optimization Problem}

Let us first denote by $\mathbf{u}' \in \mathbb{R}^n$ the  set of new scores derived from equation (\ref{f:L3}), i.e. 
\[ \forall u \in \mathbf{u} \text{, } u' = u + \alpha \log(p) + \beta(\mathbf{x}, m) \log(m/p) \] 

From a ranking perspective, $\mathbf{u}'$ is one permutation of $\mathbf{u}$ in the $\binom{n}{2} = n (n-1) / 2$ permutation space of $n$ items, and to measure the distance between our two score vectors in this space we can make use of the Kendall tau correlation measure~\cite{kendall1938new} that counts the number $n_c$ of concordant pairs versus the number $n_d$ of discordant pairs between the two vectors. Thus given:
%i.e. $n_c(\mathbf{u}, \mathbf{u}') = \sum_{i<j} \I(r(u_i) > r(u_j)) \I(r(u'_i) > r(u'_j)) + \I(r(u_i) < r(u_j)) \I(r(u'_i) < r(u'_j))$ and $n_d(\mathbf{u}, \mathbf{u}') = \sum_{i<j} \I(r(u_i) > r(u_j)) \I(r(u'_i) < r(u'_j)) + \I(r(u_i) < r(u_j)) \I(r(u'_i) > r(u'_j))$.
\begin{align*}
n_c(\mathbf{u}, \mathbf{u}') =& \sum_{i<j} \I(r(u_i) > r(u_j)) \I(r(u'_i) > r(u'_j)) \\
& + \I(r(u_i) < r(u_j)) \I(r(u'_i) < r(u'_j)) 
\end{align*}
\begin{align*}
n_d(\mathbf{u}, \mathbf{u}') =& \sum_{i<j} \I(r(u_i) > r(u_j)) \I(r(u'_i) < r(u'_j)) \\
& + \I(r(u_i) < r(u_j)) \I(r(u'_i) > r(u'_j))  
\end{align*}
The Kendall tau correlation measure can then be defined as:
\begin{align}
\label{f:K}
K(\mathbf{u}, \mathbf{u}') &=  \frac{n_c(\mathbf{u}, \mathbf{u}') - n_d(\mathbf{u}, \mathbf{u}')}{n (n-1) / 2}  
\end{align}
which has been shown recently to be a positive-definite kernel~\cite{jiao2016kendall}. This means that we can make use of the kernel trick~\cite{hofmann2008kernel} and optimize over this metric in our optimization problem. More precisely, we follow \cite{jiao2016kendall} by defining the mapping function $\phi: \mathbb{R}^n \rightarrow \mathbb{R}^{n (n-1) / 2} $,  as well as its smooth version $\tilde{\phi}$ which is given by transforming the indicator function $\I$ of the function $\phi$ to its sigmoid counterpart $\sigma(x) = \frac{1}{1+\exp(- \theta x )}$:
\begin{align*}
\phi(\mathbf{u}) &= \left( \frac{1}{\sqrt{n (n-1) / 2}} ( \I(r(u_i) > r(u_j)) - \I(r(u_i) < r(u_j)) ) \right)_{i<j} \\
\tilde \phi(\mathbf{u}) &= \left( \frac{1}{\sqrt{n (n-1) / 2}} ( \sigma(u_i - u_j) - \sigma(u_j - u_i) ) \right)_{i<j} 
\end{align*}
from which we can define the kernelized version of Kendall tau as:
\begin{align}
\hat K(\mathbf{u}, \mathbf{u}') = \phi(\mathbf{u})^{\T} \tilde \phi(\mathbf{u}')
\end{align}
where we make use of the smooth $\tilde{\phi}$ on the new score $\mathbf{u}'$ since learning will occur on this side of the kernel. 

We are now armed to present our novel optimization problem. As already stated, the task is two-fold: first, we want to learn a new ranking order $\mathbf{u}'$ of the items, described by features $\mathbf{X}$, which maximizes the NDCG of the intermediary's commissions $\mathbf{m}$. This can be expressed as a standard learning-to-rank problem similar to equation (\ref{f:Lr}), but defined here as $\mathcal{L}_r(\mathbf{m} | \mathbf{X})$  and optimized for the new scores $\mathbf{u}'$. Second, we also want to minimize the distance of the new ranking $\mathbf{u}'$  to the original ranking  $\mathbf{u}$, which we write down using our Kendall tau metric $K(\mathbf{u}, \mathbf{u}')$. Our full optimization problem can be then written as: 
\begin{align}
\label{best_b}
&\min_{\beta} \mathcal{L}(\mathbf{m} | \mathbf{u}, \mathbf{X} ) = \mathcal{L}_r(\mathbf{m} | \mathbf{X}) + \gamma (1-\hat K(\mathbf{u}, \mathbf{u'})) %\\
%&\text{where } K(\mathbf{u}, \mathbf{u}') = \phi(\mathbf{u})^{\T} \tilde \phi(\mathbf{u}') \nonumber
\end{align} 

In problem (\ref{best_b}), we can see the kernelized Kendall tau metric $\hat K(\mathbf{u}, \mathbf{u}')$ as playing the role of a similarity-based regularizer with the original ranking order $\mathbf{u}$ being the reference point to the new ranking order $\mathbf{u}'$. Also, the hyper parameter $\gamma$ gives the balance between the two terms of our  optimization problem; with a high $\gamma$, the new ranking order $\mathbf{u}'$ will not diverge too much from the original one, but neither will the commission, while with a low $\gamma$ items with a high commission will be more likely pushed on top, increasing then the chance of gaining more profit for the intermediary, unless they have a low utility (consumer preference) score. 
%In terms of training, we    makes use of the smooth mapping function $\tilde{\phi}$ on the score vector $\mathbf{u}'$ because we want to solve our problem according to the new ranking with the original score vector $\mathbf{u}$ kept fixed. 

Overall, problem (\ref{best_b}) can be seen as a reformulation of the multi-objective problem (\ref{f:L3}) in the learning-to-rank framework; it finds exactly the best weight $\beta$ that maximize the intermediary's commission with the constraint of maintaining the importance of customer preferences $\mathbf{u}$. 
Therefore by solving problem (\ref{best_b}) we also solve problem (\ref{f:L3}) which is exactly what we want.  

\subsection{Training}

We will see now how to train our function $\beta$ for solving problem (\ref{best_b}). For a given item $i$, the derivative of the loss with respect to weight $\beta_i$ is given by: \[\frac{\delta \mathcal{L}}{\delta \beta_i} = \frac{\delta \mathcal{L}_r}{\delta u'_i} \frac{\delta u'_i}{\delta \beta_i} - \lambda \frac{\delta K}{\delta u'_i} \frac{\delta u'_i}{\delta \beta_i} \] 

Because both the kernelized Kendall tau and the NDCG surrogate loss work in the $\binom{n}{2}$ permutation space, we can combine the derivatives of these two functions in a pairwise learning approach. For the NDCG surrogate loss $\mathcal{L}_r$, we follow the LambdaRank approach of \cite{burges2010ranknet} which decomposes this pairwise loss by first defining the so-called ``lambda'' gradient: \[ \lambda_{ij} = \frac{-\theta}{1+\exp(\theta (u'_i - u'_j))} |\Delta_{ij}^ \text{NDCG}|\] 
Then we have: 
\begin{align}
\frac{\delta \mathcal{L}_r}{\delta \beta_i} =   \left[ \sum_{j:m_i > m_j}^n \lambda_{ij} - \sum_{j:m_j > m_i}^n \lambda_{ji} \right] \log(m_i / p_i)
\end{align}
While for the  kernelized Kendall tau $K$ this can be decomposed by item pairs so that we have: 
\begin{align}
\frac{\delta K}{\delta \beta_i} =  \frac{1}{\sqrt{n (n-1) / 2}}  \sum_j^n \phi(\mathbf{u})_{ij} \left[ \sigma'(u'_i - u'_j) + \sigma'(u'_j - u'_i) \right] \log(m_i / p_i)
\end{align}
where $\sigma'$ is the derivative of the sigmoid function. 
The overall training complexity of our approach is thus the same as any pairwise learning to rank algorithm with $O(n^2)$. 

%% file: experiment.tex
\section{Experimental Evaluation}

%\begin{table}
%\caption{NDCG@10 Results on Expedia Hotel Searches}
%\label{tab:ndcg}
%\begin{tabular}{|c|c|c|}
%\hline
%& Clicks+Bookings & Margin  \\ \hline
%Expedia sort & 0.3987 & 0.7646 \\ \hline  
%Baseline & 0.3641 & 0.8922 \\ \hline
%LRR  & 0.3752 & 0.8922 \\ \hline
%\end{tabular}
%\end{table}

\begin{table}
\caption{Relative Improvement of NDCG@10 on Expedia Hotel Searches versus Expedia sort}
\label{tab:ndcg}
\begin{tabular}{|c|c|c|}
\hline
& Clicks+Bookings & Margin  \\ \hline
LS & -8.7\% & 16.7\% \\ \hline
LRR  & -5.9\% & 16.7\% \\ \hline
\end{tabular}
\end{table}

\begin{table}
\caption{Risk-sensitive Results on Expedia Hotel Searches}
\label{tab:risk}
\begin{tabular}{|c|c|c|}
\hline
& Clicks+Bookings  & Margin \\
\hline
Risk & 16.2\% & 36.7\% \\
\hline
Reward & 20.9\% & 55.8\% \\
\hline
\end{tabular}
\end{table}

We have evaluated our approach on an in-house Expedia dataset built from world-wide hotel searches collected during 2016. The dataset describes pairs of search queries and hotels, where the hotels are ranked according to the Expedia sort algorithm and associated with either clicks or bookings implicitly provided by the consumers. 
It actually contains 1.5M search queries, each associated with 30 hotels in average among which at least one is booked. 
Our feature space $\mathcal{X}$ is composed of various search and hotel attributes, but for the purpose of learning the weights $\beta$ we did not include the hotel price $p$ since it is a component of the margin $m$, which is our target attribute in problem (\ref{best_b}).   In these experiments we use the matrix factorization technique \cite{Abernethy2006} for the underlying recommendation algorithm. Note also that because of lack of time we did not use a validation set and all the reported metrics here are from the training set. This is a clear weakness of our experiments that we will correct in the near future.

For the baseline model, we used the approach described in \cite{krasnodebski2016}.
This approach models the weights $\alpha$ and $\beta$ of problem (\ref{f:L3}) using an additional sigmoid function. %to model the relation between the margin percent and price components. 
The optimal weights are found using a line search to minimize the weighted aggregate Kendall tau distances between our sort model and target rankings of our different objective criteria (e.g. the original CVR-optimized hotel rankings and an `ideal' ranking specification for margin defined by business rules). Note that this model has been thoroughly A/B tested and thus represents a challenging baseline. In the followings, we will call this line-search approach by LS.

For evaluation, we used a number of metrics. First, we used the NDCG@10 metric to evaluate the ranking of the three different scoring methods; the original Expedia sort (based purely on $u_i$), the LS model described above and our learning-to-re-rank (LRR) approach, relatively to the two objective criteria (\ref{f:L3}); the customer clicks+bookings and the intermediary margin. The results are provided in Table \ref{tab:ndcg}. We can see there that both LS and our LRR approach deliver similar performances compared to the original sort scores;  the two methods give a lift of +16.7\% on the NDCG of margin, but at the same time this incurs a decrease of -8.7\% and -5.9\%  in terms of the NDCG of customer preferences.

We suspected that the similar  performances of Table \ref{tab:ndcg} between LS and our LRR method are due to an averaging effect of the NDCG metric over all the queries. Thus, we used as second metric a risk-sensitive metric that basically counts the percent of queries $q$ where our LRR method beats (respectively, loses against) the baseline model for a given NDCG metric:     
\begin{align*}
\text{Risk} &= \frac{1}{|q|} \sum_{q} \text{NDCG@10(LRR)} < \text{NDCG@10(Baseline)} \\ 
\text{Reward} &= \frac{1}{|q|} \sum_{q} \text{NDCG@10(LRR)} > \text{NDCG@10(Baseline)} 
\end{align*}
The results are provided in Table \ref{tab:risk}. We can see there that the risk of LRR relatively to LS is systematically lower than its reward on both clicks+bookings and margin. In other words, our LRR method beats the LS baseline more times than it loses, where we obtain the highest reward (more than 50\% of queries) on the margin objective, i.e. in principle, LRR should achieve more revenue per transaction than the current LS method while keeping CVR at the same level or better.

%% file: conclusion.tex
\section{Conclusion and Future Work}

In this work, we have proposed a novel learning-to-re-rank approach for solving multi-objective recommendation problems involving multiple stakeholders. 
We have addressed the difficult task of learning a linear combination of potentially conflicting stakeholder objectives by defining a novel learning to (re-)rank optimization problem built on the kernel version of the Kendall tau correlation metric~\cite{jiao2016kendall}. 
 The experimental results on an Expedia dataset suggest that our approach is effective in solving this problem; it delivers the best performance trade-off for the two objectives under consideration; CVR and margin, compared to the original single-objective (CVR) model and baseline multi-objective model. 
 
As future work, we plan to A/B test the real-world performance of our model against the current baseline to see if it is truly effective in increasing both revenue and CVR. We also aim to extend our approach to additional objectives than the three ones described in section \ref{sec:prob}.For instance, we would also  like to consider various supplier participation metrics in the Expedia market place (cancellations, price competitiveness and room availabilities, etc.), which are important business indicators for Expedia in order to stay competitive against other online travel agencies.